\documentclass[12pt]{article}
\usepackage[english]{babel}
\usepackage{amsmath,amssymb,amsthm}
\usepackage{epsfig,float}
\usepackage{ifpdf}
\usepackage{empheq}					% left brace on a subequations environment

\ifpdf %-- si pdflatex
\usepackage[pdftex,colorlinks=true,urlcolor=blue]{hyperref}	% plus joli sur écran (colorie les liens au lieu d'utiliser des boites)
\graphicspath{{pdf/}}
\DeclareGraphicsExtensions{.pdf,.png,.jpg}
\hypersetup{
  pdfsubject={},
  pdfkeywords={reaction diffusion},
  bookmarks=false,
  pdfcreator=pdflatex
}
\else %-- si latex
\usepackage[colorlinks=true,urlcolor=blue]{hyperref}	% plus joli sur écran (colorie les liens au lieu d'utiliser des boites)
\graphicspath{{eps/}}
\DeclareGraphicsExtensions{.eps,.ps}
\hypersetup{
  pdfsubject={},
  pdfkeywords={reaction diffusion},
  bookmarks=false,
}
\fi

\makeatletter
\AtBeginDocument{
  \hypersetup{
    pdftitle = {\@title},
    pdfauthor = {\@author}
  }
}
\makeatother

\newcommand{\BE}{\begin{equation}}
\newcommand{\EE}{\end{equation}}
\newcommand{\be}{\begin{equation*}}
\newcommand{\ee}{\end{equation*}}
\newcommand{\R}{\mathbb{R}}

\newtheoremstyle{mon_style}		%Name
{3pt}					%Space above
{3pt}					%Space below
{}					%Body font
{}					%Indent amount 1
{\bf}					%Theorem head font
{:}					%Punctuation after theorem head
{.5em}					%Space after theorem head 2
{}					%Theorem head spec (can be left empty, meaning ‘normal’)
\theoremstyle{mon_style}
\newtheorem{De}{Definition}
\newtheorem{theo}{Theorem}
\newtheorem{conj}[theo]{Conjecture}

\title{Generalized traveling wave in heterogeneous media: \\
Allee effect can inhibit invasion}
\author{Benoit Sarels \\
\small{benoit.sarels@cmla.ens-cachan.fr} \\
\small{Centre de math\'ematiques et de leurs applications} \\
\small{\'Ecole normale sup\'erieure de Cachan \& CNRS, UMR 8536} \\
\small{61, avenue du Pr\'esident Wilson} \\
\small{94235 Cachan cedex, France}
}
\date{October 31, 2012}

\begin{document}
\maketitle
\begin{abstract}
In this paper, we study a simple one-dimensional model of reaction-diffusion with bistable non-linearity and in heterogeneous media.
The bistable term accounts for the so-called Allee effect, 
and the heterogeneity in the media is localized.
We recall the definition of a transition wave used in similar situations by well-known authors, 
and propose an alternative definition for discussion.
We call it generalized traveling wave.
As a consequence, we give new results of pinning in such media, due to Allee effect.
\end{abstract}
\tableofcontents
\newpage

%%%%%%%% 1
\section{Introduction}
%%%%%%%%
\subsection{Framework}
We will focus in this paper on the following reaction-diffusion equation:
\BE \label{eq:RD_rx_fu}
\left\{
\begin{aligned}
\partial_t u& = \Delta u + r(x)~f(u) \\
u(\cdot,0)&=u_0
\end{aligned}
\right.
\EE
It models a wide range of phenomena in biology, which will be our main concern in applications.
While the homogeneous version of this equation ($r(x)\equiv r_0$) has been well studied since the pioneering works of Fisher \cite{fisher} and Kolmogorov and al. \cite{kpp}, 
the heterogeneous ones are an actual field of research.
We assume here that heterogeneity occurs only in the reactivity of the equation, through the $r(x)$ function, while the diffusivity is constant ($d=1$ can be achieved by a convenient rescaling).

The most part of the paper is devoted to the so-called bistable reaction term:
\BE \label{eq:bistable}
f(u)=u(1-u)(u-a)
\EE
with $0<a<1$.
It is even convenient to suppose $0<a<\frac{1}{2}$, as we will see below.
Whenever possible, we will discuss the extension of our results to the monostable case (Fisher--KPP)
\BE \label{eq:monostable}
f(u)=u(1-u)
\EE

Our interest here is in equations that are local perturbations of the homogeneous one: 
we suppose then that $r(x)=r_0+r_1(x)$, with $r_0>0$, $r_1$ compactly supported, continuous, and so that $r(x)$ remains non-negative: $||r_1||_{\infty} \leq r_0$.

While it is very common to see applications to biology make use of a monostable reaction term, 
let us recall that in a number of cases it is not a reasonable assumption near $0$.
In our mind, the reaction term describes the growth of a biological quantity (be it a population density, a concentration, a proportion of a population sharing a given attribute, $\cdots$).
A logistic growth is the first improvement over an exponential one ($f(u)=r u$), taking account of the existence of the habitat or environment (with capacities, resources, limits).
Under such an hypothesis, the growth is no more proportional to $u$, and saturates when the population reaches the so-called carrying capacity of the habitat $K>0$ ($f(u)= r u (1-\frac{u}{K})$).

The next improvement concerning the reaction term is related to the Allee effect.
The Allee effect is the correlation between individual fitness and global density.
At low densities, the survival of an individual can be made more complicated by increased predation or difficulties in mating.
In this kind of situation, it it not reasonable to have $f$ logistic at low densities.
We can differentiate between weak Allee effect (for which $f'(0)=0$ is a good approximation), 
and Strong Allee effect ($f'(0)<0$ and thus $f<0$ near zero).
The bistable term models then a population growth with a strong Allee effect.
Such a case is also called the "heterozygote inferior" in a genetic context \cite{aronson}.

The reactivity $r$ takes account of the intensity of the reaction, here intensity of the growth (population case), of the selection (genetic case).
%%%%%%%%
\subsection{Classical theory}
We recall here quickly the known results for equation \eqref{eq:RD_rx_fu} when $r$ is constant equal to $1$ and $f$ bistable:
\BE \label{eq:RD_1_bistable}
\left\{
\begin{aligned}
\partial_t u& = \Delta u + u(1-u)(u-a) \\
u(\cdot,0)&=u_0
\end{aligned}
\right.
\EE

We are interested here in classical traveling waves of this equation: $u(x,t)=U(x-ct)$.
For clarity, we choose to investigate waves going to the right, with invasion of the state $0$ by the state $1$.
Such a $U$ is solution of
\BE
U''+cU'+U(1-U)(U-a) \quad U(-\infty)=1 \quad U(+\infty)=0
\EE
It is known \cite{aronson,fife} that $U$ is unique up to translation, and $c$ is related to $a$:
\BE \label{eq:sol_prog}
U(z)=\frac{1}{1 + \exp { \left( z \right)}} \quad c= \sqrt{\frac{1}{2}}(1-2a) % \quad w=\sqrt{\frac{2}{r}}
\EE
We see that the speed $c$ is positive as long as $a<\frac{1}{2}$.
%If $a>\frac{1}{2}$, the speed is negative.
This is related to the sign of the integral of $f(u)$ on $[0,1]$.
For a constant positive $r$ we have
\BE \label{eq:sol_prog_r}
U(z)=\frac{1}{1 + \exp { \left( z \sqrt{\frac{r}{2}} \right)}} \quad c= \sqrt{\frac{r}{2}}(1-2a) % \quad w=\sqrt{\frac{2}{r}}
\EE
A comment that can be done here is the $r$-dependence on the speed and profiles of the wave.
Formulas in \eqref{eq:sol_prog_r} show that bigger $r$ leads to quicker and steeper wave.
Thus we can say that when $r$ is big, the media is favorable for the invasion by the state $1$ and the invasion is sudden.
When $r$ is low, the media is less favorable, and the invasion is progressive.

%%%%%%%% 2
\section{Heterogeneous media: transition waves vs. generalized traveling waves}
\subsection{Transition waves}
More general media can exhibit range more or less favorable for propagation.
As explained before, we study here media that are local perturbation of the homogeneous case.
This allows the relaxation of the initial condition to the traveling wave, due to convergence results (see for example \cite{fife}).
Then, from an initial condition far enough from zero, 
the solution is near the classical traveling wave before it reaches the zone where $r$ varies.

We expect that in the bistable case, stability of the leading edge of the wave will lead to similar behaviors when $r$ varies.
This has been reinforced by numerical simulations made in \cite{pre11}.

To investigate such situations, one can use the notion of transition fronts defined for example in \cite{ber_ham}
\footnotemark\footnotetext{
It seemed to us during our bibliographic research that these authors previously used the term of generalized traveling wave, before they coined the term transition front (or transition wave).
They seem now to stick to the latter.
}.
A transition front for equation \eqref{eq:RD_rx_fu} is a global in time solution that connects the two steady states:
\be
u(-\infty,t)=1 \quad u(+\infty,t)=0
\ee
for each $t$, and whose $\varepsilon$-width is uniformly bounded in time for all $\varepsilon$.
The $\varepsilon$-width is defined as follow:
$\mbox{diam} \{x\in \R | u(x,t) \geq \varepsilon \mbox{ and } 1-u(x,t) \geq \varepsilon\}$

This has been done in \cite{ber_ham_mata,nolen}.

%%%%%%%%
\subsection{Generalized traveling waves}
We propose here another approach, already developed in \cite{pre11}, though we did not state there an appropriate definition.
The main idea is to follow the front at each time (each time step for the numerics).
We then get another type of result: an approximation result instead of a convergence one.

Our work is backed upon an extensive numerical study, 
which gave birth to a conjecture.
This conjecture reads: every reasonable initial datum will evolve following the scheme
\be
u(x,t) \approx U \left( \frac{x-x_0(t)}{w(t)} \right)
\ee
where $U$ is the solution of the homogeneous equation \eqref{eq:RD_1_bistable}, $x_0$ the center (or position) of the wave, $w$ its width.
This width is defined in the same way as above: 
$\mbox{diam} \{x\in \R | u(x,t) \geq \delta \mbox{ and } 1-u(x,t) \geq \delta\}$ 
for an appropriate $\delta$ chosen once for all.
The profile $U$ makes it unnecessary to control the width for any $\delta$, 
for widths for different thresholds are related with a fixed coefficient.
%$\approx$ means that we control the norm $L^\infty$ of the difference of the two sides.

\begin{conj}
~\\
Let $u$ be the solution of the Cauchy problem \eqref{eq:RD_rx_fu} with reaction term \eqref{eq:bistable}.\\
Let $U$ be the wave profile in homogeneous medium defined in \eqref{eq:sol_prog}.\\
Then there exist $x_0$ and $w$ such as $u(x,t) - U \left( \frac{x-x_0(t)}{w(t)} \right)$ remains small in $L^\infty$-norm provided as $r$ varies few enough:
\be
\forall \varepsilon >0\: \exists \eta >0\: \quad\mbox{such as}\quad \forall t >0\: \exists (x_0(t),w(t))
\ee
\be
||r_1||_{\infty} \leq \eta \Rightarrow ||u(\cdot,t) - U \left( \frac{\cdot-x_0(t)}{w(t)} \right)||_{\infty} \leq \varepsilon
\ee
\end{conj}
\begin{De}
~\\
Such a solution of \eqref{eq:RD_rx_fu}-\eqref{eq:bistable} is called a generalized traveling wave
\footnotemark\footnotetext{
We use this term because we believe it is left free by the authors previously mentioned.
}
.
\end{De}
Full proof of this conjecture will be left for a future paper.
%We can indicate to the reader that the proof is based on an inertial manifold method.
We want to emphasize here that this definition covers cases of defects narrow or wide, symmetric or antisymmetric as defined in \cite{pre11}.
The big defects though are not covered, but they behave in a similar manner.

If this conjecture holds, and we strongly believe it to do so, it is an interesting result because we have ways to compute $x_0$ and $w$.
This is dealt with in the next section.

Note also that this conjecture does not hold for the Fisher--KPP case due to the instability of the state $0$.
Instead, secondary outbreaks occur in advance of the wave \cite{pre11}.
%%%%%%%% 3
\section{Velocity and width}
%%%%%%%%%%
\subsection{Moments}
Given that $U(z)=\frac{1}{1+\exp z}$, we have $U'(z)=-\frac{\exp z}{(1+\exp z)^2}$.
Thus $U'$ is fastly decreasing at infinity and its moments are well defined.
The three first ones are:
\BE \label{eq:moments}
\left\{
\begin{aligned}
\alpha_0 &= \int_{-\infty}^{+\infty} U'(z)~dz = -1 \\
\alpha_1 &= \int_{-\infty}^{+\infty} z U'(z)~dz = 0 \\
\alpha_2 &= \int_{-\infty}^{+\infty} z^2 U'(z)~dz = -\frac{\pi^2}{3}
\end{aligned}
\right.
\EE
If we assume very roughly that we have
\be
u(x,t)=U\left( \frac{x-x_0(t)}{w(t)}\right)
\ee
then we can differentiate $u$:
\be
\partial_x u(x,t)=\frac{1}{w(t)} U'\left( \frac{x-x_0(t)}{w(t)}\right)
\ee
Making, for a given $t$ the change of variable $z=\frac{x-x_0(t)}{w(t)}$ in the three integrals of \eqref{eq:moments}, yields
\BE \label{eq:moments_resolus}
\left\{
\begin{aligned}
1 &=-\int_{-\infty}^{+\infty} \partial_x u(x,t) ~dx \\
x_0 &=-\int_{-\infty}^{+\infty} x \partial_x u(x,t) ~dx \\
w^2 &= -\frac{3}{\pi^2} \left( \int_{-\infty}^{+\infty} x^2 \partial_x u(x,t) ~dx +x_0^2 \right)
\end{aligned}
\right.
\EE
So that we have formulas for $x_0$ and $w$.
This first way to obtain them is the least interesting, 
because it requires to compute (numerically) the solution $u$, then its partial derivative, and some additional integrals.

%%%%%%%%%%
\subsection{Least square fit}
The second way to obtain the position and width of the generalized traveling wave has been developed in \cite{pre11}.
It consists in fitting (at each time) the solution $u$ with an appropriate profile.
We have consistently checked that this fitting procedure can be done with a numerical routine providing a very small error (in $L^\infty$-norm).
This evaluation of $x_0$ and $w$ is excellent, and takes less time than the one above.

%%%%%%%%%%
\subsection{Balance laws}
A much better solution is given by the following system \eqref{eq:dx0dw_bistable}.
This system comes from two simple balance laws from the original partial differential equation \eqref{eq:RD_rx_fu},
and has been derived in \cite{pre11}.
It is a basic system of two ordinary differential equations that yield the evolution of the position and width:
\BE \label{eq:dx0dw_bistable}
\left\{
\begin{aligned}
\dot{x_0} &= w \int_{-\infty}^{+\infty} r(wz+x_0)f(U(z)) ~dz \\
\dot{w}   &= \frac{1}{3w} + w \int_{-\infty}^{+\infty} r(wz+x_0) (1-2U(z)) f(U(z)) ~dz
\end{aligned}
\right.
\EE
By this way, evaluation of the position and width is simpler and quicker.
%%%%%%%% 4
\section{Applications}

We derive in this section some easy consequences of the equations \eqref{eq:dx0dw_bistable}.
These equations have the great benefit to link the position and width of the generalized traveling wave to the heterogeneity $r$.
They show that the qualitative behavior of the front around an obstacle is not trivial (see \cite{pre11} for details).
We have notably a slowing down of the front before it reaches any positive defect.
We add here two more pinning criteria to the one previously given in this article.
%%%%%%%%%%
\subsection{\label{critere_2}Pinning of the front by an antisymmetric defect}
Let us consider the case of a narrow antisymmetric defect, modeled by an Heaviside step function:
\be
r(x)=r_0+r_1~H(x)
\ee
where $H(x)=0$ if $x$ is negative or superior to a given constant $R$ big enough, and $H(x)=1$ between $0$ and $R$.
We deal with the case $r_1>0$.
\begin{enumerate}
\item {\bf Equations for the critical width \\}
The system of equations \eqref{eq:dx0dw_bistable} can be rewritten for such a defect $r(x)$, and when the front stops we have:
\be
\dot{x_0}=0,\:\dot{w}=0
\ee
we get:
\be
\left\{
\begin{aligned}
		&{}r_0 w \left(\frac{1-2a}{2}\right)	&+ &{}r_1 w \int_{\frac{-x_0}{w}}^{+\infty} f(U(z))~dz 		&=0 \\
\frac{1}{3w}- 	&{}r_0 \frac{w}{6} 			&+ &{}r_1 w \int_{\frac{-x_0}{w}}^{+\infty} (1-2U(z))f(U(z))~dz	&=0
\end{aligned} 
\right.
\ee
or:
\begin{subequations}\label{grp_critere_2}
\begin{empheq}[left=\empheqlbrace]{align}
		&{}r_0 w \left(\frac{1-2a}{2}\right)	&+ &{}r_1 w S_1 \left(\frac{-x_0}{w} \right)	&=0 \label{first_critere_2} \\
\frac{1}{3w}- 	&{}r_0 \frac{w}{6} 			&+ &{}r_1 w S_2 \left(\frac{-x_0}{w} \right)	&=0 \label{second_critere_2}
\end{empheq}
\end{subequations}
We can first simplify the first equation by $w$,
then we note that the left part of the left hand side of \eqref{first_critere_2} is always positive.
In the right hand side, we supposed $r_1>0$.
To balance the equation, it is necessary that $S_1 \left(\frac{-x_0}{w}\right)$ contributes negatively.
This is indeed possible given the plot of $S_1(y) = \frac{1-2a(1+e^y)}{2(1+e^y)^2}$ : $S_1$ has a negative minimum (see figure \ref{plot_S1}).
\begin{figure}[H]
\centerline{
\epsfig{file=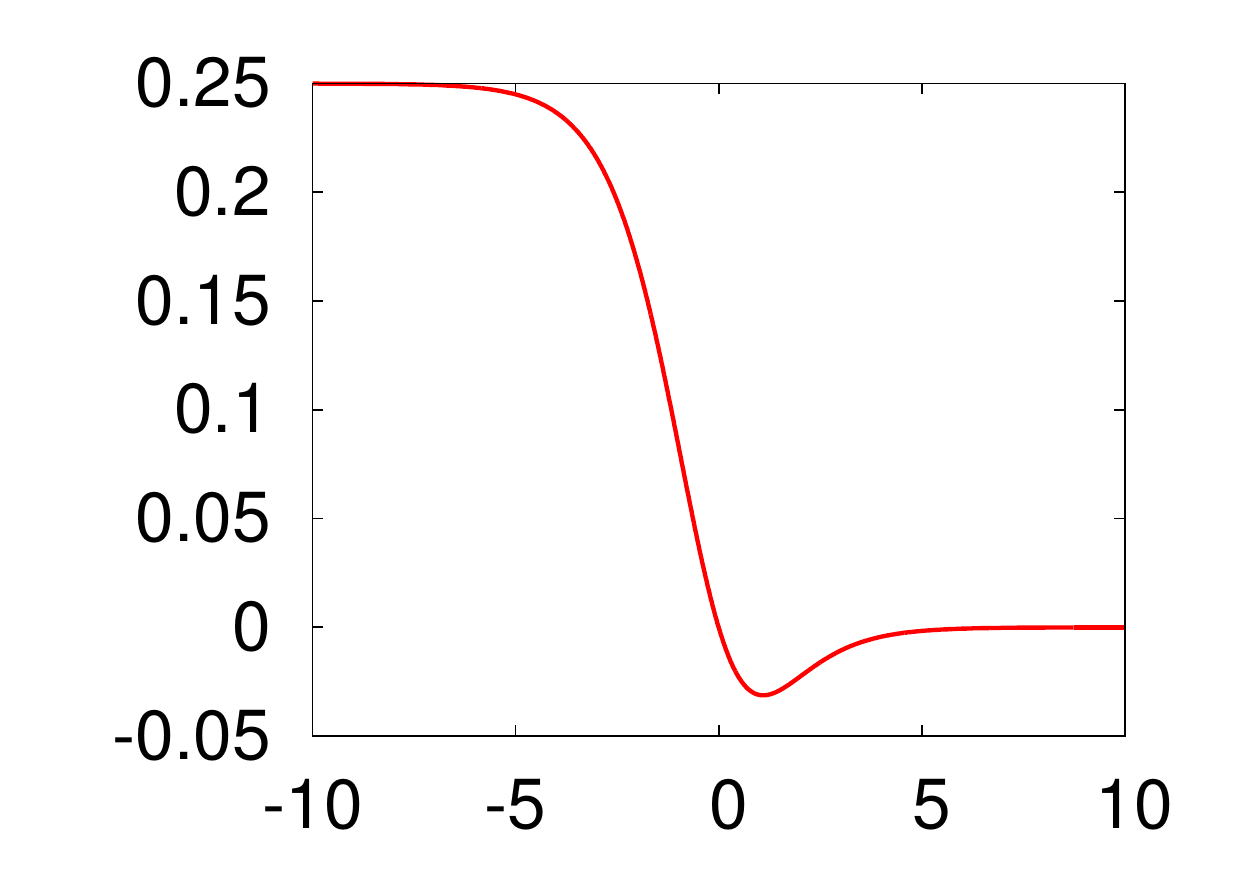,width=0.5\linewidth}
}
\caption{Plot of $S_1$. We chose $a=0.25$.}
\label{plot_S1}
\end{figure}
The front stops if and only if:
\be
r_0 \left(\frac{1-2a}{2} \right) < -r_1 \min (S_1)
\ee
For the critical case, we have equality:
\BE\label{hauteur_critique_Heaviside}
r_0 \left(\frac{1-2a}{2} \right) = -r_{1c} \min (S_1)
\EE
This shows the existence of a threshold over which pinning occurs, and under which the wave goes on propagating.
Note that this pinning is counter-intuitive for $r_1>0$, given the considerations before given on the $r$-dependence of the speed.

\item {\bf Resolution of the equations \\}

The minimum of $S_1$ and its argument $z_{min}$ can be obtained because $S_1'(y)=-R_1(y)$ where $R_1(y)=f(U(y))= \frac{e^y(1-a-a e^y)}{(1+e^y)^3}$.
We have $R_1(y)=0 \Leftrightarrow 1-a-ae^y=0$, so
\be
z_{min} = \ln{\left( \frac{1-a}{a} \right)}
\ee
then
\be
\min (S_1) = S_1(z_{min})= \frac{1-2a(1+\frac{1-a}{a})}{2(1+\frac{1-a}{a})}=- \frac{a^2}{2}
\ee
This gives the critical height thanks to \eqref{hauteur_critique_Heaviside} :
\be
r_{1c}=\frac{r_0(1-2a)}{a^2}
\ee

We find next the width $w$ at pinning thanks to \eqref{second}, with the substitution of $\frac{-x_0}{w}$ by $\ln{\left( \frac{1-a}{a} \right)}$ and of $r_1$ by the obtained quantity.
After a little algebra, we get :
\be
w=\sqrt{\frac{2}{r_0}}\frac{1}{\sqrt{1+(1-2a)(3-2a)}}
\ee
and finally :
\be
x_0=-\sqrt{\frac{2}{r_0}}\frac{\ln{\left( \frac{1-a}{a} \right)}}{\sqrt{1+(1-2a)(3-2a)}}
\ee
\end{enumerate}

We have evaluated this result using the following numerical application with $a=0,3$ and $r_0=1$: % a comparer a une simulation
\be
\left\{
\begin{aligned}
r_{1c} &=\frac{40}{9} \approx 4.44 \\
w &= \sqrt{\frac{50}{49}} \approx 1.01 \\
x_0 &= - \sqrt{\frac{50}{49}} \ln{\left(\frac{7}{3}\right)} \approx -0.86
\end{aligned}
\right.
\ee
which compares very well to the values coming from the simulations (error of $3 \%$).
%%%%%%%%%%
\subsection{\label{critere_3}Pinning of the front by a no-reaction zone} % c'est dans la thèse
Let us consider the case of a no-reaction zone, modelized by the following function:
\be
r(x)=
\left\{
\begin{aligned}
r_0 \quad \mbox{if}& \quad x<0\\
0 \quad \mbox{if}& \quad 0\leq x\leq x_1\\
r_0 \quad \mbox{if}& \quad x>x_1.
\end{aligned}
\right.
\ee
We ask the question of the minimal width required to stop the front.
We call this minimal width $x_{1c}$, so that we have the following dichotomy:
\begin{itemize}
\item if $x_1<x_{1c}$, then the front is not stopped by the obstacle. We will say that the front goes through the obstacle, it goes on propagating after the zone.
\item if $x_1 \geq x_{1c}$, then the front is stopped, pinned in the no-reaction zone.
The wave is not moving any longer, neither in one direction nor in the other: its speed of spread decreases to zero.
When it reaches the no-reaction zone, the front slows down and changes shape.
At the limit, the front tends to an affine profile (solution of the heat equation with Dirichlet boundary conditions).
\end{itemize}

\begin{figure}[H]
\centerline{
\epsfig{file=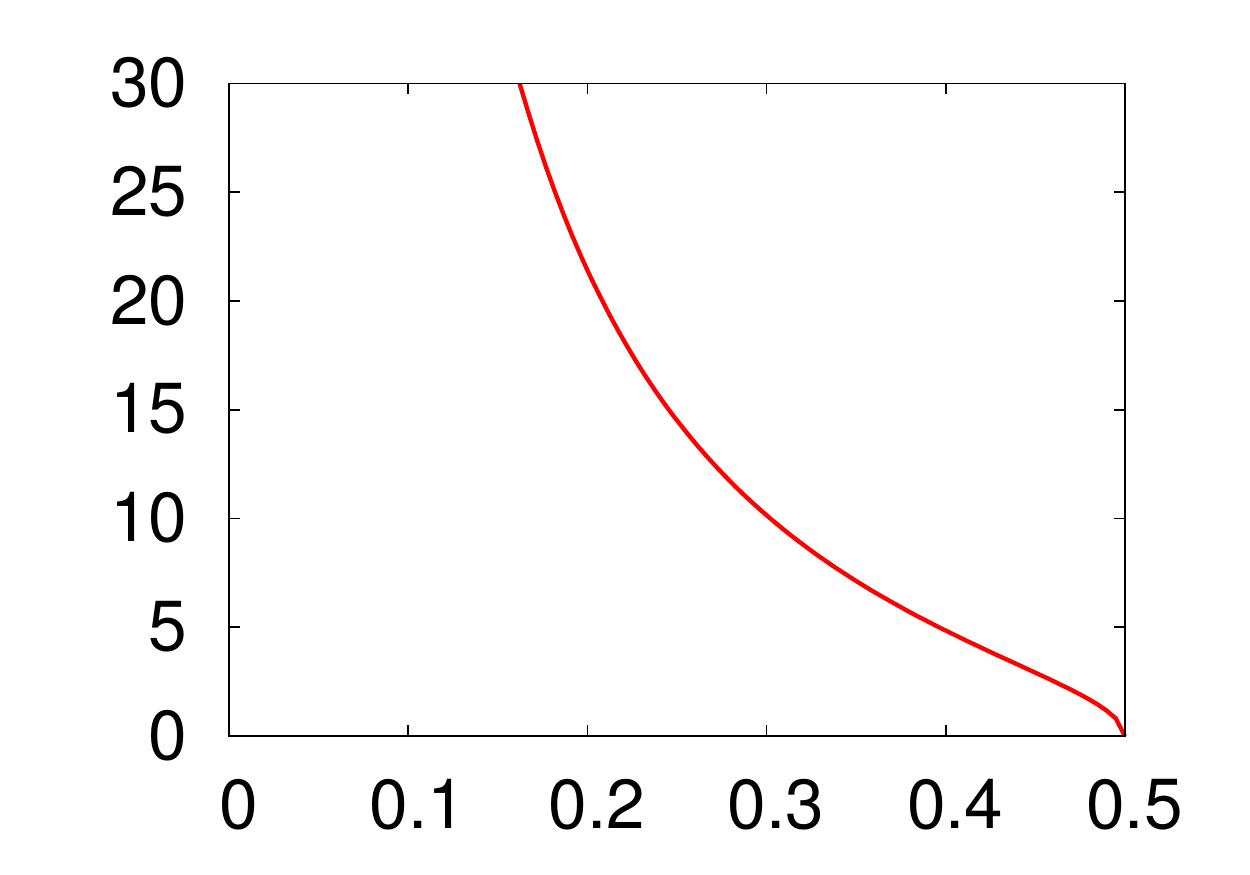,width=0.5\linewidth}
}
\caption{Plot of 
$\sqrt{\frac{(b-a)^3}{-1+3a-8a^3+9a^4+7a^4b-2ab+b-3a^5-2a^3b}} \left( \ln{\left( \frac{1-a}{a} \right)} -\ln{\left( - \frac{a+a^2-1+b}{2a+a^2-1} \right)} \right)$}
\label{fa_pinning_3}
\end{figure}

We solve this problem in a straight-forward manner:
\begin{enumerate}
\item {\bf Equations for the critical width \\}
The system of equations \eqref{eq:dx0dw_bistable} can be rewritten for such a defect $r(x)$, and when the front stops we have:
\be
\dot{x_0}=0,\:\dot{w}=0
\ee
Then we get:
\be
\left\{
\begin{aligned}
&w~r_0 \left( \int_{-\infty}^{\frac{-x_0}{w}} f(U(z))~dz + \int_{\frac{x_{1c}-x_0}{w}}^{+\infty} f(U(z))~dz \right) =\: 0 \\
&\frac{1}{3w} + w~r_0 \left( \int_{-\infty}^{\frac{-x_0}{w}} (1-2U(z))f(U(z))~dz + \int_{\frac{x_{1c}-x_0}{w}}^{+\infty} (1-2U(z))f(U(z))~dz \right) =\: 0,
\end{aligned}
\right.
\ee
or stated otherwise:
\begin{subequations}\label{grp}
\begin{empheq}[left=\empheqlbrace]{align}
%\left\{\begin{aligned}
%\begin{align}
&w~r_0 \left( T_1 \left(\frac{-x_0}{w} \right) + S_1 \left(\frac{x_{1c}-x_0}{w} \right) \right) =\: 0 \label{first}\\
&\frac{1}{3w} + w~r_0 \left( T_2 \left(\frac{-x_0}{w} \right) + S_2 \left(\frac{x_{1c}-x_0}{w} \right) \right) =\: 0. \label{second}
%\end{align}
%\end{aligned}\right.
\end{empheq}
\end{subequations}
The first equation can be divided by $w~r_0$ (positive quantity), and the wave stops when both terms in the left hand side of \eqref{first} balance each other.
The integrals can be calculated, we have:
\be
T_1(y) = \frac{e^y(2+e^y-2a(1+e^y))}{2(1+e^y)^2}
\ee
It is obvious that $T_1(y) = \frac{e^y(1+(1-2a)(1+e^y))}{2(1+e^y)^2}$ is always positive for all $a \in [0,\frac{1}{2}]$.
We show in figure \ref{plot_T1} an example of such a plot. % for $a=0.3$ check !!??
\begin{figure}[H]
\centerline{
\epsfig{file=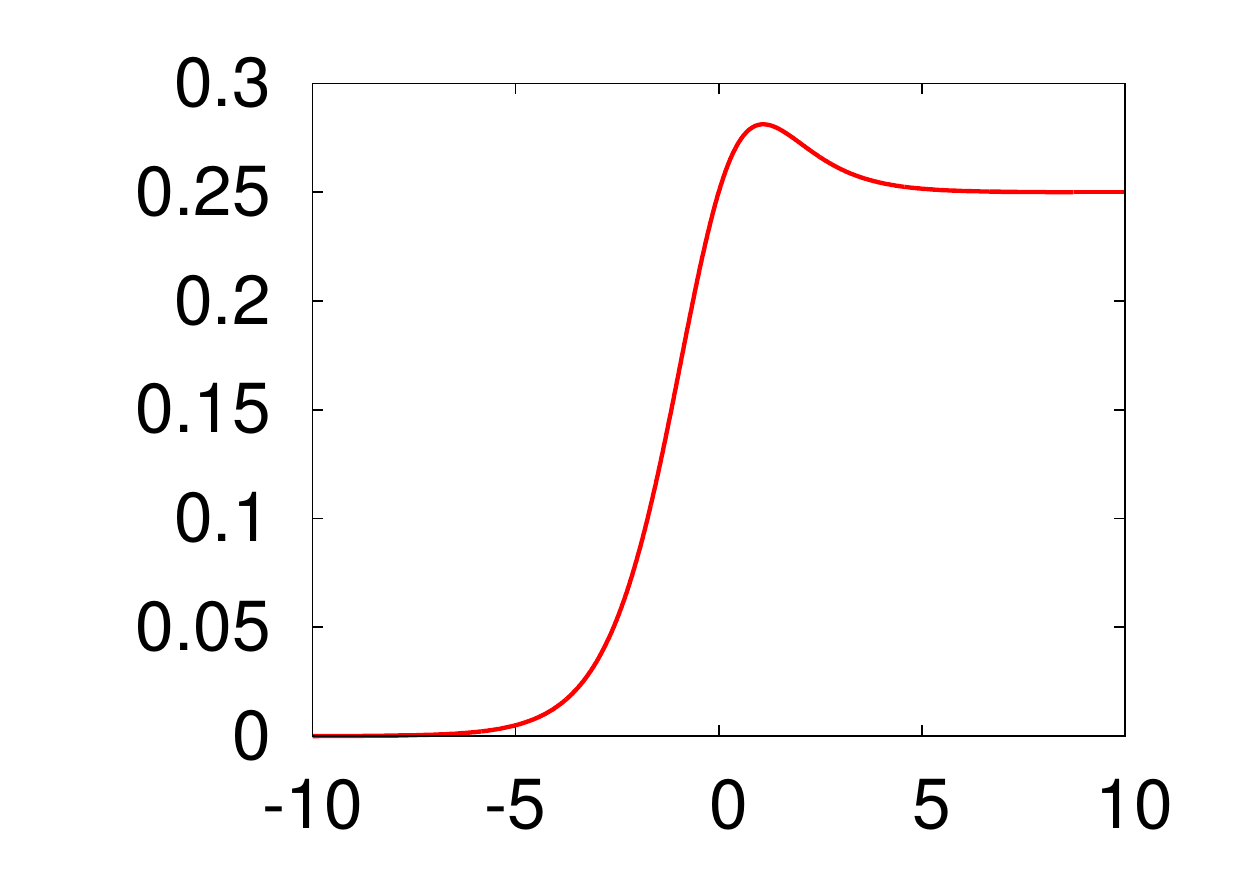,width=0.5\linewidth}
}
\caption{Plot of $T_1$. We chose $a=0.25$.}
\label{plot_T1}
\end{figure}
So that the equation can be verified, it is necessary that $S_1 \left(\frac{x_{1c}-x_0}{w} \right)$ be negative.
This is indeed possible given the profile of $S_1$ (see figure \ref{plot_S1}).
$S_1$ has a strictly negative minimum, but for widths below the critical width, this negative part is not enough to stop the front.
For the critical width $x_{1c}$, we have exactly
\BE \label{balance}
T_1 \left(\frac{-x_0}{w} \right) = - \min S_1
\EE

\item {\bf Values of $\frac{-x_0}{w}$ and $\frac{x_{1c}-x_0}{w}$ when the wave stops \\}
We have seen previously that the minimum of $S_1$ and its argument $z_{min}$ can be easily computed:
\begin{align*}
z_{min}  = {} & \ln{\left( \frac{1-a}{a} \right)} \\
\min (S_1) = {} & S_1(z_{min})= - \frac{1}{2(1+\frac{1-a}{a})^2}.
\end{align*}
We have then
\be
\frac{x_{1c}-x_0}{w} = z_{min} = \ln{\left( \frac{1-a}{a} \right)}
\ee
and we get the expression, as a function of $a$ and $b=\sqrt{1-2a}$, of the ratio of both collective coordinates when the front stops for the critical width:
\be
\frac{x_0}{w}=-\ln{\left( - \frac{a+a^2-1+b}{2a+a^2-1} \right)},
\ee
thanks to equation \eqref{balance}, because $-\min (S_1)$ has a unique antecedent by $T_1$.

\item {\bf Resolution of the equations \\}
We replace the quantities $\frac{-x_0}{w}$ and $\frac{x_{1c}-x_0}{w}$ by their values in equation \eqref{second},
we can then solve and we obtain the width $w_c$ of the front when it stops:
\be
w_c = \frac{1}{\sqrt{2r_0}} \sqrt{\frac{(b-a)^3}{-1+3a-8a^3+9a^4+7a^4b-2ab+b-3a^5-2a^3b}},
\ee
and the critical width $x_{1c}$ as well.
The pinning criteria writes:
\BE
\label{critere_arret_3}
x_1 \geqslant x_{1c} = w_c \left( \ln{\left( \frac{1-a}{a} \right)} -\ln{\left( - \frac{a+a^2-1+b}{2a+a^2-1} \right)} \right)
\EE
\end{enumerate}
It is interesting to note that the critical width depends on $r_0$ via the inverse of its square root.
We can check on figure \ref{fa_pinning_3} that the right hand side of \eqref{critere_arret_3} is a decreasing function of $a$ on $[0,\frac{1}{2}]$, with infinite limit at $0$ and null at $\frac{1}{2}$.
This is a proof of consistency of the standardized model, as we expect the front to be more easily pinned as $a$ approach the value $\frac{1}{2}$.

The values obtained via this criteria are in very good agreement with numerical simulations of the initial partial differential equation.
For $r_0=1$, our criteria \eqref{critere_arret_3} gives $x_{1c}\approx\frac{10.15}{\sqrt{2}}\approx7.18$ (we recall that we fixed $a=0.3$).
Numerically, we find that the front stops from $x_{1c}\approx6.90$, which yields an error inferior to $4\%$.

%%%%%%%%
\section{Conclusion}
In this paper, we proposed a notion that generalizes the classical traveling wave in homogeneous media to heterogeneous ones.
It uses two moments of the derivative of the profile $U$, which has the appropriate behavior at infinity for doing so.

We have quickly shown in this paper the power of this new tool.
Once written the evolution equations of the position and width, it becomes possible to tackle down difficult problems while answering questions that are of crucial interest to biologists.
The idea of following the front at each time is fruitful for people interested in applications, and complements the convergence results obtained in a traditional manner.


\begin{thebibliography}{99}

\bibitem{zeldovich}
Ya.B. Zeldovich and D.A. Frank-Kamenetsky (1938),
``K teorii ravnomernogo rasprostraneniya plameni'',
Dokladi Akademii Nauk SSSR, 19(9):693--697.

\bibitem{fisher}
R.A. Fisher (1937),
``The wave of advance of advantageous genes'',
Annals of Eugenics, 7, 355--369.

\bibitem{kpp}
A.N. Kolmogorov, I.G. Petrovskii and N.S. Piskunov (1937), 
``A study of the equation of diffusion with increase in the quantity of matter, and its application to a biological problem'', 
Bjul. Moskovskogo Gos. Univ. 1:7, 1 26.

\bibitem{aronson}
D.G. Aronson and H.F. Weinberger (1975),
``Nonlinear diffusion in population genetics, combustion and nerve propagation'',
in Partial Differential Equations and Related Topics, ed. J.A. Goldstein.
Lecture Notes in Mathematics 446, 5--49 New York: Springer.

\bibitem{fife}
P.C. Fife and J. McLeod (1977),
``The approach of solutions of nonlinear diffusion equations to traveling front solutions'',
Arch Ration Mech Anal 65:335--361.

%\bibitem{PhysRevE.84.041108}
\bibitem{pre11}
J.G. Caputo and B. Sarels (2011),
``Reaction-diffusion front crossing a local defect'',
Phys. Rev. E, volume 84, issue 4, pages 041108.
%  numpages = {8},
%  month = {Oct},
%  doi = {10.1103/PhysRevE.84.041108},
%  url = {http://link.aps.org/doi/10.1103/PhysRevE.84.041108},
%  publisher = {American Physical Society}

\bibitem{ber_ham}
H. Berestycki and F. Hamel (2007), 
``Generalized traveling waves for reaction-diffusion equations'',
in H. Brezis (Ed.), Perspectives in Nonlinear Partial Differential Equations,
in Contemp. Math., vol. 446, Amer. Math. Soc.

\bibitem{ber_ham_mata}
H. Berestycki, F. Hamel, and H. Matano (2009),
``Bistable traveling waves around an obstacle'',
Commun. Pure Appl. Math. 62, 729--788

\bibitem{nolen}
J. Nolen, J.M. Roquejoffre, L. Ryzhik and A. Zlatos (2012), % Zlatoš
``Existence and Non-Existence of Fisher--KPP Transition Fronts'',
Arch. Rational Mech. Anal. 203 217--246

\end{thebibliography}
\end{document}